\documentclass[conference]{IEEEtran}

\usepackage{graphicx}
\usepackage{epstopdf}
\usepackage{standalone}
\usepackage[nolist ]{acronym}
\usepackage{tcolorbox}

\usepackage{pdfpages}
\usepackage{tikz}
\usetikzlibrary{positioning}
\usepackage[bookmarks=false]{hyperref}
\usepackage{pdfcomment}
\usepackage{hologo}

\usepackage{xstring}

\usepackage[utf8]{inputenc}
\usepackage{marvosym}

\usepackage{algorithm}
\usepackage{algpseudocode}
\usepackage{tikz}

\usepackage{listings}
\definecolor{ListingBackground}{rgb}{0.97,0.97,0.97}
\usepackage{pgfplots}
\pgfplotsset{compat=newest}
\pgfplotsset{
    box plot/.style={
        /pgfplots/.cd,
        fill=blue!30,
        only marks,
        mark=-,
        mark size=0.2em,
        /pgfplots/error bars/.cd,
        y dir=plus,
        y explicit,
    },
    box plot box/.style={
        /pgfplots/error bars/draw error bar/.code 2 args={%
            \draw  ##1 -- ++(.2em,0pt) |- ##2 -- ++(-.2em,0pt) |- ##1 -- cycle;
        },
        /pgfplots/table/.cd,
        y index=2,
        y error expr={\thisrowno{3}-\thisrowno{2}},
        /pgfplots/box plot
    },
    box plot top whisker/.style={
        /pgfplots/error bars/draw error bar/.code 2 args={%
            \pgfkeysgetvalue{/pgfplots/error bars/error mark}%
            {\pgfplotserrorbarsmark}%
            \pgfkeysgetvalue{/pgfplots/error bars/error mark options}%
            {\pgfplotserrorbarsmarkopts}%
            \path ##1 -- ##2;
        },
        /pgfplots/table/.cd,
        y index=4,
        y error expr={\thisrowno{2}-\thisrowno{4}},
        /pgfplots/box plot
    },
    box plot bottom whisker/.style={
        /pgfplots/error bars/draw error bar/.code 2 args={%
            \pgfkeysgetvalue{/pgfplots/error bars/error mark}%
            {\pgfplotserrorbarsmark}%
            \pgfkeysgetvalue{/pgfplots/error bars/error mark options}%
            {\pgfplotserrorbarsmarkopts}%
            \path ##1 -- ##2;
        },
        /pgfplots/table/.cd,
        y index=5,
        y error expr={\thisrowno{3}-\thisrowno{5}},
        /pgfplots/box plot
    },
    box plot median/.style={
        /pgfplots/box plot
    },
    boxplot/every median/.style={
    	ultra thick,dashed,cyan
    }
}

\definecolor{flexicolor}{RGB}{46,49,146}
\definecolor{amaricolor}{RGB}{237,28,36}

\usepackage{xspace}

\usepackage[binary-units=true]{siunitx}
\usepackage{psfrag}
\usepackage{graphicx}
\usepackage{tabularx,booktabs}
\usepackage{multirow}
\usepackage{rotating}

\renewcommand{\baselinestretch}{1}

\usepackage{multirow}

\usepackage[cmex10]{amsmath}
\usepackage[caption=false,font=footnotesize]{subfig}
\usepackage{stfloats}
\newif\ifdoubleblind 

\begin{document}

\newcommand{\paperTitle}{Template}
\newcommand{\paperAuthors}{Benjamin Sliwa, Manuel Patchou, and Christian Wietfeld}
\newcommand{\paperEmails}{$\{$Benjamin.Sliwa, Manuel.Mbankeu, Christian.Wietfeld$\}$@tu-dortmund.de}

\newcommand{\sfw}{0.24}
\newcommand{\figurePadding}{0pt}
\newcommand{\figureTopPadding}{\figurePadding}
\newcommand{\figureBottomPadding}{\figurePadding}
\newcommand{\red}[1]{\colorbox{red}{\textbf{TODO}: #1}}

\newcommand{\dummy}[3]
{
	\begin{figure}[b!]  
		\begin{tikzpicture}
		\node[draw,minimum height=6cm,minimum width=\columnwidth]{\LARGE #1};
		\end{tikzpicture}
		\caption{#2}
		\label{#3}
	\end{figure}
}

\newcommand{\wDummy}[3]
{
	\begin{figure*}[b!]  
		\begin{tikzpicture}
		\node[draw,minimum height=6cm,minimum width=\textwidth]{\LARGE #1};
		\end{tikzpicture}
		\caption{#2}
		\label{#3}
	\end{figure*}
}

\newcommand{\basicFig}[7]
{
	\begin{figure}[#1]  	
		\vspace{#6}
		\centering		  
		\includegraphics[width=#7\columnwidth]{#2}
		\caption{#3}
		\label{#4}
		\vspace{#5}	
	\end{figure}
}
\newcommand{\fig}[4]{\basicFig{#1}{#2}{#3}{#4}{0cm}{0cm}{1}}

\newcommand{\subfig}[3]
{%
	\subfloat[#3]%
	{%
		\includegraphics[width=#2\textwidth]{#1}%
	}%
	\hfill%
}

\newcommand\circled[1] 
{
	\tikz[baseline=(char.base)]
	{
		\node[shape=circle,draw,inner sep=1pt] (char) {#1};
	}\xspace
}
\begin{acronym}
	\acro{LIMoSim}{Lightweight ICT-centric Mobility Simulation}
	\acro{DDNS}{Data-driven Network Simulation}
	\acro{DES}{Discrete Event Simulation}
	\acro{ns-3}{Network Simulator 3}
	\acro{WEKA}{Waikato Environment for Knowledge Analysis}
	\acro{LIMITS}{Lightweight Machine Learning for IoT Systems}
	\acro{MNO}{Mobile Network Operator}
	\acro{LTE}{Long Term Evolution}
	\acro{eNB}{evolved Node B}
	\acro{UE}{User Equipment}
	\acro{TCP}{Transmission Control Protocol}
	\acro{LENA}{LTE-EPC Network Simulator}
	\acro{ITU}{International Telecommunication Union}
	\acro{RF}{Random Forest}
	\acro{GPR}{Gaussian Process Regression}
	\acro{RSRP}{Reference Signal Received Power}
	\acro{RSRQ}{Reference Signal Received Quality}
	\acro{SINR}{Signal-to-interference-plus-noise Ratio}
	\acro{CQI}{Channel Quality Indicator}
	\acro{TA}{Timing Advance}
	\acro{RMSE}{Root Mean Square Error}
	\acro{MAE}{Mean Absolute Error}
	\acro{OSM}{OpenStreetMap}
	\acro{SUMO}{Simulation of Urban Mobility}
	\acro{CAT}{Channel-aware Transmission}
	\acro{ML-CAT}{Machine Learning CAT}
	\acro{5GAA}{5G Automotive Association}
	\acro{QoS}{Quality of Service}
	\acro{REM}{Radio Environmental Map}
	\acro{KPI}{Key Performance Indicator}
	\acro{PRB}{Physical Resource Block}
\end{acronym}

\acresetall
\title{The Best of Both Worlds: Hybrid Data-Driven and Model-Based Vehicular Network Simulation}

\ifdoubleblind
\author{\IEEEauthorblockN{\textbf{Anonymous Authors}}
	\IEEEauthorblockA{Anonymous Institutions\\
		e-mail: Anonymous Emails}}
\else
\author{\IEEEauthorblockN{\textbf{\paperAuthors}}
	\IEEEauthorblockA{Communication Networks Institute,	TU Dortmund University, 44227 Dortmund, Germany\\
		e-mail: \paperEmails}}
\fi

\maketitle

\begin{tikzpicture}[remember picture, overlay]
\node[below=5mm of current page.north, text width=20cm,font=\sffamily\footnotesize,align=center] {Accepted for presentation in: 2020 IEEE Global Communications Conference (GLOBECOM)\vspace{0.3cm}\\\pdfcomment[color=yellow,icon=Note]{
@InProceedings\{Sliwa2020the,\\
	Author = \{Benjamin Sliwa and Manuel Patchou and Christian Wietfeld\},\\
	Title = \{The Best of Both Worlds: Hybrid Data-Driven and Model-Based Vehicular Network Simulation\},\\
	Booktitle = \{2020 IEEE Global Communications Conference (GLOBECOM)\},\\
	Year = \{2020\},\\
	Address = \{Taipei, Taiwan\},\\
	Month = \{Dec\}\\
\}
}};
\node[above=5mm of current page.south, text width=15cm,font=\sffamily\footnotesize] {2020~IEEE. Personal use of this material is permitted. Permission from IEEE must be obtained for all other uses, including reprinting/republishing this material for advertising or promotional purposes, collecting new collected works for resale or redistribution to servers or lists, or reuse of any copyrighted component of this work in other works.};
\end{tikzpicture}

\begin{abstract}

%
%
The analysis of the end-to-end behavior of novel mobile communication methods in concrete evaluation scenarios frequently results in a methodological dilemma: Real world measurement campaigns are highly time-consuming and lack of a controllable environment, the derivation of analytical models is often not possible due to the immense system complexity, system-level network simulations imply simplifications that result in significant derivations to the real world observations.
%
%
In this paper, we present a hybrid simulation approach which brings together model-based mobility simulation, multi-dimensional \acp{REM} for efficient maintenance of radio propagation data, and \ac{DDNS} for fast and accurate analysis of the end-to-end behavior of mobile networks.
For the validation, we analyze an opportunistic vehicular data transfer use-case and compare the proposed method to real world measurements and a corresponding simulation setup in \ac{ns-3}. In comparison to the latter, the proposed method is not only able to better mimic the real world behavior, it also achieves a $\sim$300 times higher computational efficiency.

\end{abstract}

\IEEEpeerreviewmaketitle

\section{Introduction}

%
%
Anticipatory communication \cite{Bui/etal/2017a} has emerged as a novel networking paradigm focusing on \emph{context-aware} optimization of decision processes in highly dynamic wireless communication systems such as vehicular networks.
In a recent report \cite{5GAA/2020a}, the \ac{5GAA} has pointed out that \emph{predictive \ac{QoS}} -- e.g., the ability to forecast the achievable data rate along a predicted trajectory -- will be one of the key enablers for connected and automated driving.
Another recent research trend in this domain is \emph{non-cellular-centric} networking. Hereby, the mobile devices become part of the network fabric and contribute explicitly or implicitly to the overall network optimization \cite{Coll-Perales/etal/2019a}. As an example, opportunistic data transfer for delay-tolerant applications (e.g., vehicle-as-a sensor) allows to dynamically schedule data transmissions with respect to the anticipated resource efficiency \cite{Sliwa/etal/2019d}.
%
%
%

%
%
However, the development and optimization of these novel mobile networking methods confronts researchers and engineers with a \emph{methodological dilemma}. Real world experiments involve massive efforts and are impacted by an uncontrollable environment. Analytical modeling is often not possible due to the immense complexity of the evaluation scenario. System-level network simulation requires assumptions and simplifications which result in an accuracy degradation for complex real world scenarios (see Sec.~\ref{sec:related_work}).

%
%
In recent work \cite{Sliwa/Wietfeld/2019d}, we have presented \ac{DDNS} as a novel machine learning-enabled method for simulating vehicular communication networks. \ac{DDNS} learns an end-to-end model of a target \ac{KPI} in a \emph{concrete} scenario based on empirical measurements. The learned model can then be utilized for the performance evaluation of novel methods under study. However, since \ac{DDNS} relies on replaying real world network conditions as context \emph{traces}, it is bound to the trajectories of the measurements and does not allow to modify the mobility behavior of the vehicles.

%
%
In this paper, we bring together the key features of \ac{DDNS} with \emph{model-based} mobility simulation in order to benefit from the best of both worlds. For this purpose, we decouple the \ac{DDNS} method from the trace-based approach through usage of multi-dimensional \acp{REM}.

%
%
The remainder of the paper is structured as follows. After discussing related work in Sec.~\ref{sec:related_work}, we present the proposed solution approach in Sec.~\ref{sec:approach}. Afterwards, the applied methodology is introduced in Sec.~\ref{sec:methods} and finally, the results of the performance evaluation are presented and discussed in Sec.~\ref{sec:results}. The developed simulation framework and the raw results are provided in an Open Source manner\footnote{Source code available at \url{https://github.com/BenSliwa/Hybrid_DDNS}}.
\section{Related Work} \label{sec:related_work}

%
%
\textbf{Network simulation} is the de-facto standard method for analyzing the performance of mobile communication systems \cite{Cavalcanti/etal/2018a}. System-level simulations provide a controllable environment and allow to compare different methods under study in \emph{abstract} scenarios. However, the achieved results often differ significantly from real measurements in \emph{concrete} complex real world scenarios \cite{Sliwa/Wietfeld/2019d}. The major reasons for this observation are:
\emph{Simplifications} such as the usage of probabilistic shadowing models instead of explicit modeling of obstacles and materials. \emph{Assumptions} as concrete parameterizations and applied algorithms are either unknown (e.g., the traffic patterns of the cell users) or are treated confidentially by the \acp{MNO} (e.g., the applied resource schedulers and concrete parameters of the \acp{eNB}). \emph{Missing features} within the implementation of the network simulator (e.g., as discussed in Sec.~\ref{sec:methods}, \ac{CQI} and \ac{TA} are not modeled in \ac{LENA} for \ac{ns-3}).
%
%
It can be been these issues are systematically implied for the system-level network simulation method due to the need to \emph{explicitly} model and parameterize communicating entities. In contrast to that, the \ac{DDNS} method \cite{Sliwa/Wietfeld/2019d} -- which is applied in a modified version in this paper -- uses machine learning to \emph{implicitly} learn the context-dependent behavior of an observed performance indicator only based on empirical measurements.
%
%
As an alternative to model-based methods, \acp{REM} \cite{Poegel/Wolf/2015a} represent a \emph{data-driven} approach for considering radio propagation effects in wireless network simulations. Hereby, models are replaced by geospatially aggregated radio measurements which are often obtained in a \emph{crowdsensing} manner.

%
%
\textbf{Machine learning} has achieved great attention within the wireless research community \cite{Wang/etal/2020a} as its inherent capability of exposing hidden interdependencies between measurable variables allows to derive models for processes which are too complex to describe analytically.
%
%
In their technical recommendation Y.3172 \cite{ITU-T/2019a}, the \ac{ITU} presents an architectual framework for machine learning-based decision making in future networks. Hereby, a simulation-based \emph{digital twin} of the network allows to safely explore the impact of different decision alternatives before actual actions are performed in the real world \emph{underlay network}.
%
%
It can be expected that the emerging research field of machine learning-based end-to-end system modeling \cite{Doerner/etal/2018a, Sliwa/Wietfeld/2019d} will further stimulate the progression in this field.
%
%

%
%
As an example for machine learning-based radio propagation analysis, Thrane et al. \cite{Thrane/etal/2020a} propose a model-aided \emph{deep learning} method which implicitly extracts radio propagation characteristics from top-view geographical images. In comparison to ray tracing techniques which are applied in a model of the same evaluation scenario, the machine learning-enabled method is able to reduce the average \ac{RSRP} prediction error by more than 50~\%. 
However, although deep learning has achieved impressive results in the image processing domain, it is not a universal remedy for all optimization problems in engineering. In the wireless communications domain, the amount of training data is often limited since data has to be acquired in complex measurement campaigns. Due to the \emph{curse of dimensionality} \cite{Zappone/etal/2019a}, deep learning techniques often get outperformed by simpler models such as \acp{RF} \cite{Breiman/2001a} which are able to better cope with smaller data sets (e.g., for mobile data rate prediction as discussed by \cite{Sliwa/Wietfeld/2019d}).

\section{Hybrid Data-Driven and Model-Based Vehicular Network Simulation}  \label{sec:approach}

In this section, the proposed hybrid simulation method and its core modules are introduced. The overall goal is to analyze the performance of a novel \emph{method under study} in a \emph{concrete} real world scenario. 
%
%
\begin{figure}[]  	
	\vspace{0cm}
	\centering		  
	\includegraphics[width=1.0\columnwidth]{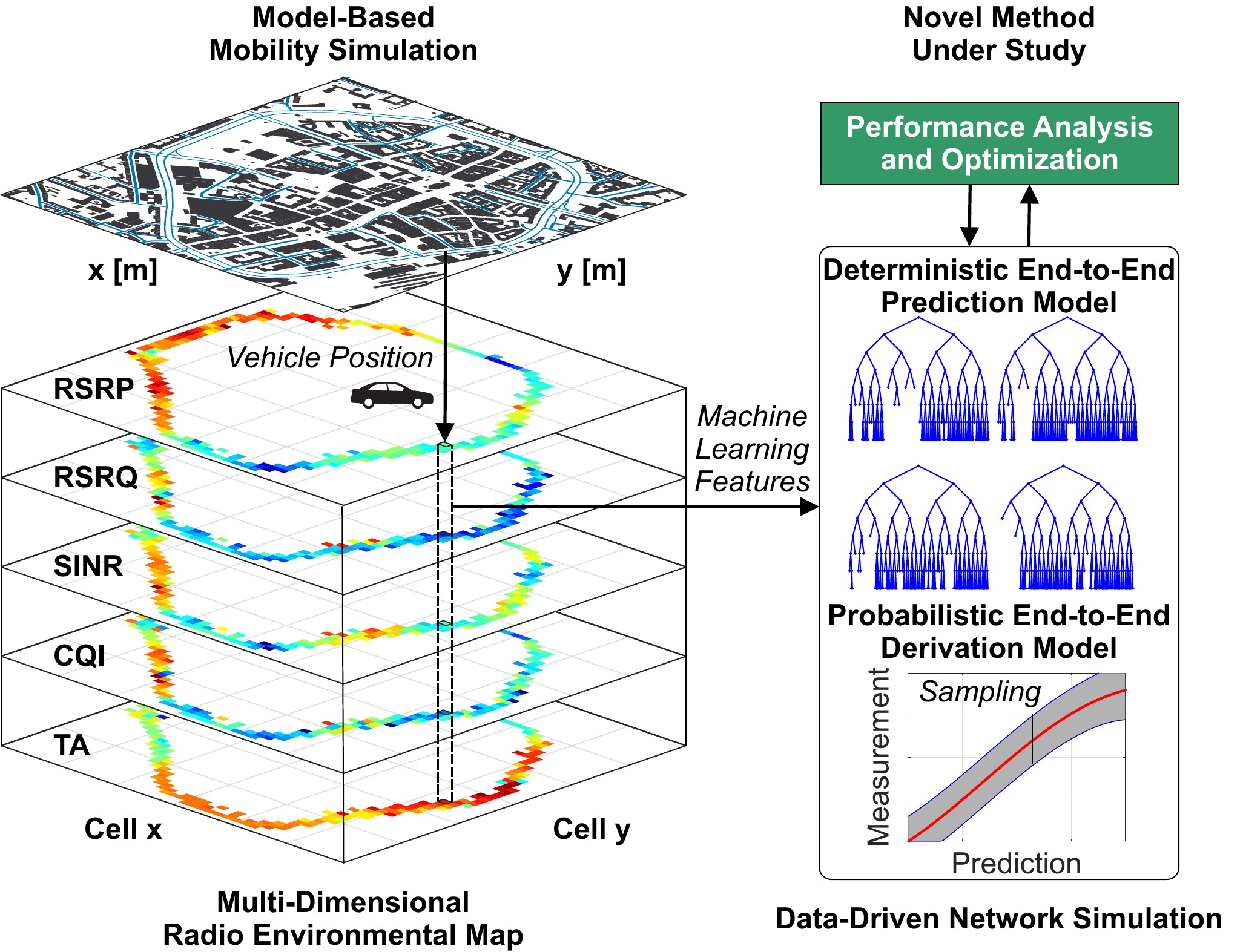}
	\vspace{-0.5cm}	
	\caption{System architecture model for the proposed hybrid vehicular network simulation method. In the offline training phase, the machine learning models utilize the whole \ac{REM} data set as a priori information. In the online application phase, predictions are performed based on the looked up values for the corresponding vehicle locations. (Map data: ©OpenStreetMap contributors, CC BY-SA).}
	\label{fig:architecture}
	\vspace{-0.5cm}	
\end{figure}
As shown in Fig.~\ref{fig:architecture}, the proposed approach consists of four core components -- the method under study, a model-based mobility simulator, a multi-dimensional \ac{REM}, and a \ac{DDNS} setup.

\textbf{Method under Study}:
In the following, we illustrate the application of the proposed method based on an example use case focusing on opportunistic vehicular sensor data transmission. For this purpose, we analyze the resulting end-to-end data rate $S$ of different transmission schemes as target \ac{KPI}.

\textbf{Model-based Mobility Simulation}:
The mobility behavior of the vehicles is represented by a mobility simulation framework which utilizes validated analytical models for the different components such as automatic cruise control and routing. Hereby, the model-based approach allows to analyze the impact of arbitrary traffic conditions and routing paths on the behavior of the method under study.
%
%
For this purpose, we apply \ac{LIMoSim} \cite{Sliwa/etal/2019c} which provides integrated support for real world map data from \ac{OSM}. 
%
%
%

\textbf{Multi-Dimensional Radio Environmental Map}: 
Within the proposed data-driven simulation approach, radio propagation and protocol effects are implicitly learned by a combination of end-to-end machine learning algorithms. For enabling this data-driven approach, it is assumed that measurement data for the target \ac{KPI} is available as \emph{a priori information}. This data can either be obtained by performing initial real world measurements, through open data sets such as \cite{Sliwa/Wietfeld/2019a, Thrane/etal/2020a}, or via crowdsensing-based services.
For predicting the end-to-end data rate $\tilde{S}$ as the considered target \ac{KPI}, the \ac{UE}-based prediction method from \cite{Sliwa/Wietfeld/2019a} is applied. Multiple features from different logical context domains are considered:
%
%
\begin{itemize}
	\item \textbf{Network context}: \ac{RSRP}, \ac{RSRQ}, \ac{SINR}, \ac{CQI}, \ac{TA}
	\item \textbf{Mobility context}: Velocity, Cell id
	\item \textbf{Application context}: Payload size of the packet
\end{itemize}
While, the features of the mobility and application domains have to be acquired online during the simulations, the network context features are maintained in a multi-dimensional \ac{REM} whereas each layer corresponds to one of the features for the machine learning process.
%
%
For a given vehicle position $\mathbf{P}(t)$, the corresponding feature set $\mathbf{\tilde{F}}(t) $ is looked up from the \ac{REM} $M$ as
%
%
\begin{equation}
	\mathbf{\tilde{F}}(t) = M \left( \lfloor \frac{\mathbf{P}(t)}{c} \rfloor \right) 
\end{equation}
with $c$ being the cell width which defines the map granularity.

\textbf{Data-driven Network Simulation}:
%
%
Finally, the end-to-end behavior of the observed \ac{KPI} is simulated based on a modified \ac{DDNS} setup. While conventional \ac{DDNS} simulations according to \cite{Sliwa/Wietfeld/2019d} are based on replaying context \emph{traces}, the proposed approach utilizes the simulated trajectories and context lookups from the \ac{REM}. \ac{DDNS} simulations rely on two main building blocks which are realized as corresponding machine learning models:
\begin{itemize}
	\item A deterministic \textbf{prediction model} is used to learn the end-to-end behavior of the considered indicator using supervised learning on the a priori data set. For the online prediction, the feature set $\mathbf{\tilde{F}}(t)$ is looked up from the \ac{REM} and the data rate $\tilde{S}(t)$ is predicted as $\tilde{S}(t) = f_{\textbf{ML}}(\tilde{F}(t))$ using the trained machine learning model $f_{\textbf{ML}}$.  Due to the findings in \cite{Sliwa/Wietfeld/2019a}, this model is represented by a \ac{RF} predictor. However, due to the deterministic nature of the learned model, identical feature sets will always result in identical predictions. In contrast to that, in the real world, the prediction models are imperfect which results in a difference between predictions and ground truth measurements.
	\item In order to represent this aspect within the simulation setup, a probabilistic \textbf{derivation model} is applied for learning the uncertainties of the prediction model of the previous step based on \ac{GPR} \cite{Rasmussen/2004a}. Hereby, the Bayesian nature of this model class is exploited, since the resulting confidence function allows to sample data values from the whole value range of a given prediction. The sampled value is then utilized as a \emph{virtual ground truth} (e.g., the achieved data rate $S(t)$ of a transmission) within the simulation setup. A visual representation of a derivation model is shown in Fig.~\ref{fig:architecture}.
\end{itemize}
For a more detailed description about the \ac{DDNS}-specific mechanisms, we forward the interested reader to \cite{Sliwa/Wietfeld/2019d}.
%
%
%

%
%
\begin{figure}[]  	
	\vspace{0cm}
	\centering		  
	\includegraphics[width=1.0\columnwidth]{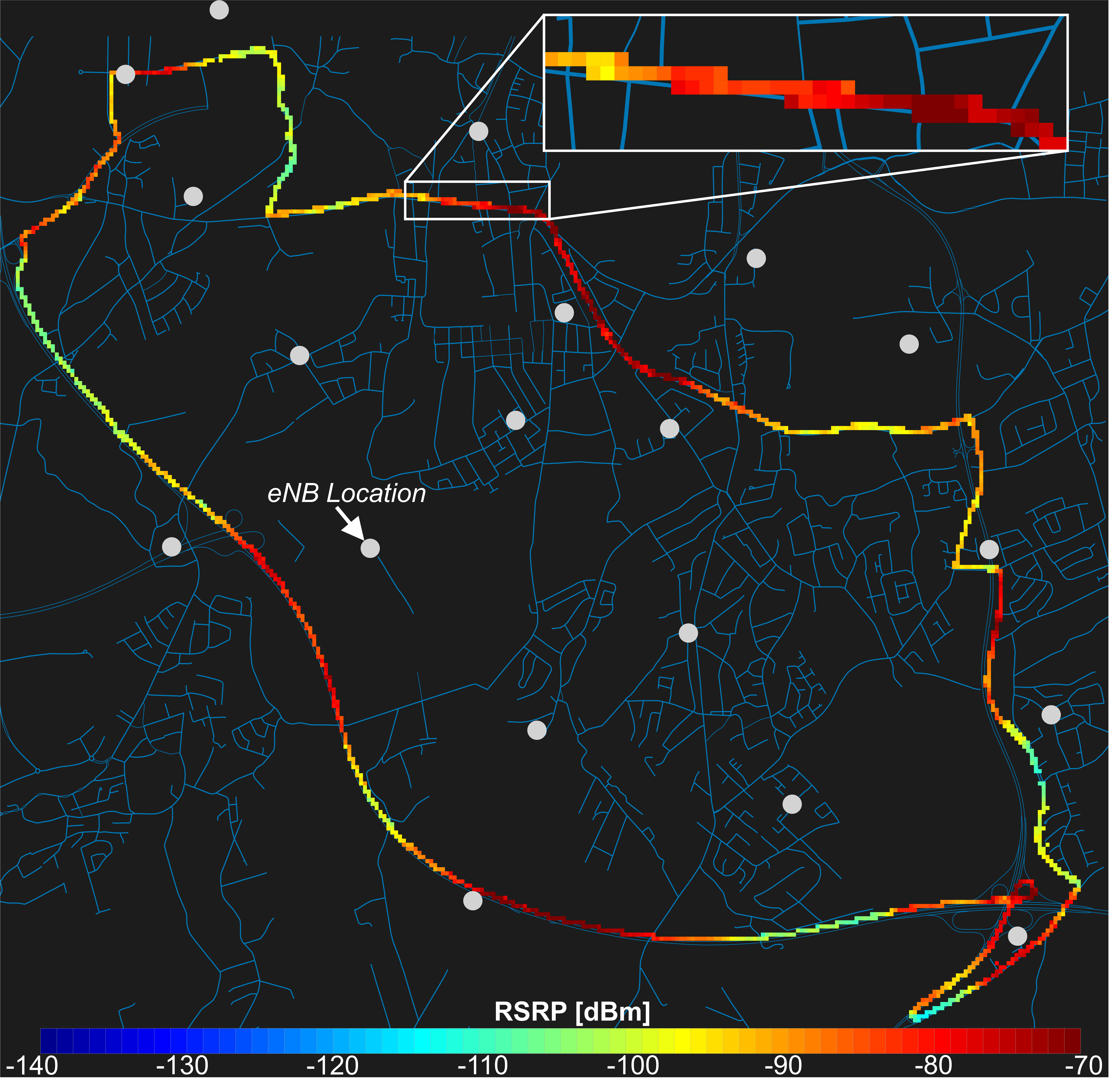}
	\vspace{-0.3cm}
	\caption{Overview about the road network topology for the evaluation scenario. The overlay shows the \ac{RSRP} layer of the \ac{REM} along the evalation track (Map data: ©OpenStreetMap contributors, CC BY-SA).}
	\label{fig:map}
	\vspace{-0.5cm}	
\end{figure}
%
%
%

%
%
\begin{figure*}[] 
	\centering
	
	\subfig{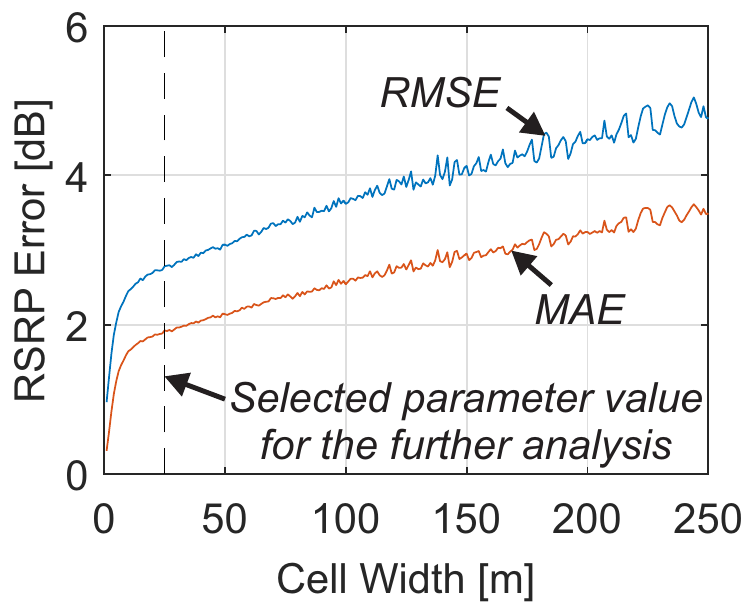}{\sfw}{}%
	\subfig{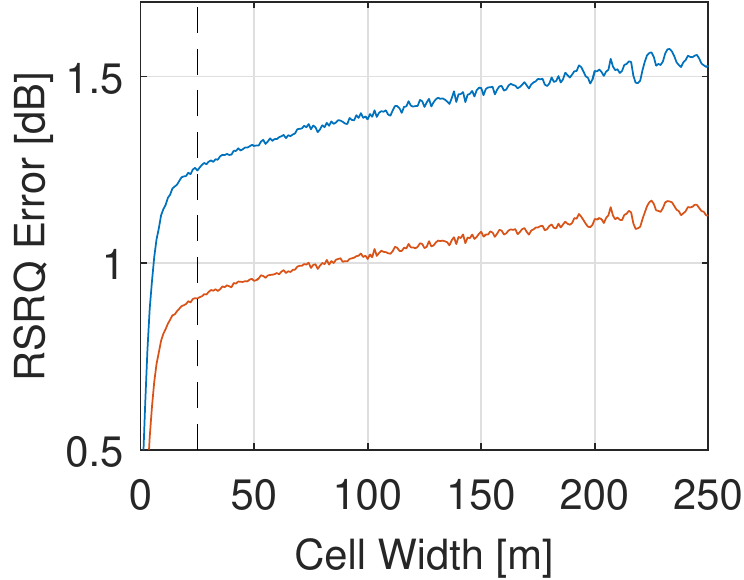}{\sfw}{}%
	\subfig{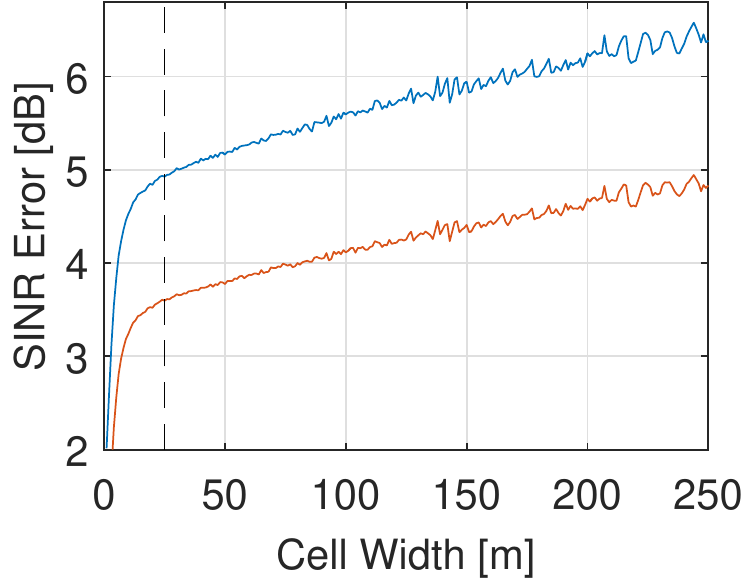}{\sfw}{}%
	\subfig{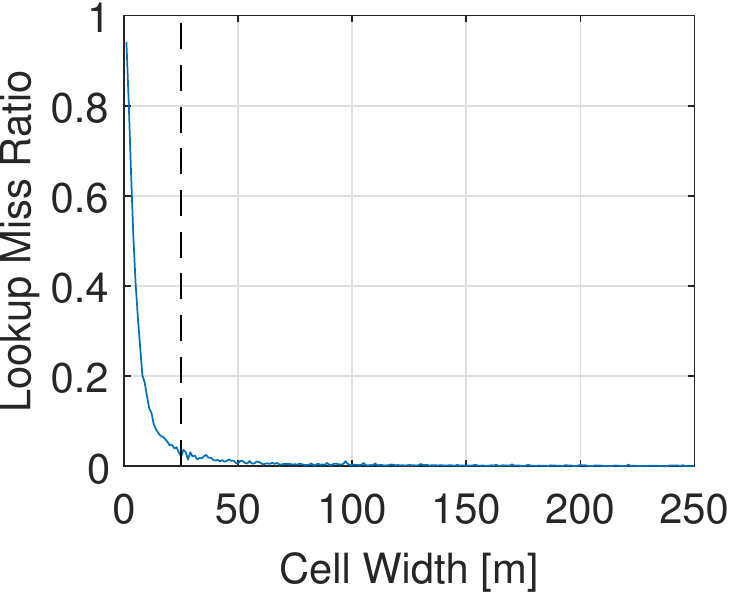}{\sfw}{}%
	
	\caption{Impact of the cell width $c$ of the radio environmental map on the resulting lookup accuracy for different network context features.}
	\vspace{-0.6cm}
	\label{fig:cm_errors}
	
\end{figure*}

%
%
\begin{figure}[] 
	\centering
	\subfig{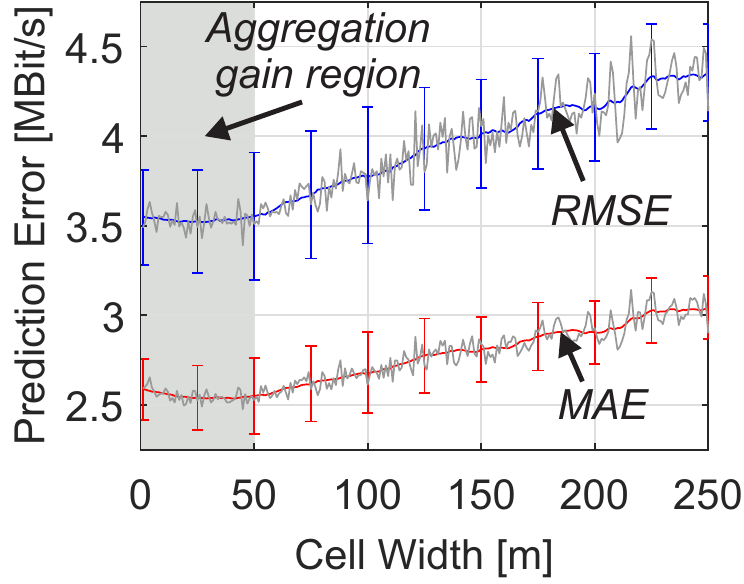}{\sfw}{Uplink}%
	\subfig{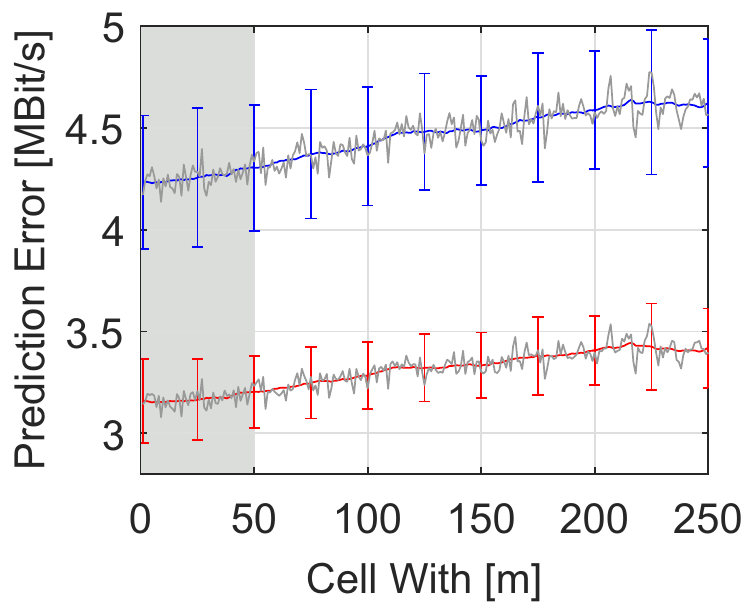}{\sfw}{Downlink}%

	\caption{Impact of the cell width of the radio environmental map on the resulting data rate prediction error. The errorbars show the standard deviation of the 10-fold cross validation.}
	\label{fig:cm_dataRate_errors}
\end{figure}

\section{Methodology}  \label{sec:methods}

In this section, the evaluation scenario as well as the tools and methods for the performance evaluation are presented. 

\subsection{Evaluation Scenario and Evaluated Methods}

For the validation of the proposed approach, we model a vehicle-as-a-sensor use case and compare the end-to-end data rates of different conventional and opportunistic data transmission schemes.
%
%
\begin{itemize}
	\item \textbf{Periodic} data transfer with a fixed interval $\Delta t=10~s$
	\item \textbf{\acf{CAT}} \cite{Ide/etal/2015a} is a probabilistic data transfer scheme which derives a transmission probability based on measurements of the current \ac{SINR}.
	\item \textbf{\acf{ML-CAT}} \cite{Sliwa/etal/2019d} is a machine-learning-based extension to \ac{CAT}. Instead of using raw network quality measurements, \ac{ML-CAT} applies an \ac{RF}-based data rate prediction which is then used to compute the transmission probability.
\end{itemize}
Data is transmitted from a moving vehicle in the uplink and downlink direction through the public cellular network using \ac{TCP}. A virtual sensor application generates 50~kByte of data per second which is buffered locally until the transmission decision is made for the whole data buffer.
Fig.~\ref{fig:map} shows the map of the evaluation scenario as well as the \ac{RSRP} layer of the \ac{REM}.

\subsection{Data Analysis}

%
%
All prediction models are trained with the Open Source \ac{LIMITS} \cite{Sliwa/etal/2020c} framework which provides high-level automation for validated \ac{WEKA} \cite{Hall/etal/2009a} models and supports the generation of \texttt{C++} code for trained machine learning models.
For the generation of the \ac{GPR}-based derivation models required for the \ac{DDNS}, we utilize the \emph{Statistics and Machine Learning Toolbox} of \texttt{MATLAB}.

%
%
As performance metrics for the resulting prediction errors, we consider \ac{RMSE} and \ac{MAE} which are computed as
%
%
\begin{equation*}
	\text{MAE} = \frac{\sum_{i=1}^{N} | \tilde{y}_{i} - y_{i}|}{N},
	\quad
	\text{RMSE} = \sqrt{\frac{\sum_{i=1}^{N} \left(  \tilde{y}_{i} - y_{i} \right)^2}{N}}.
\end{equation*}
with $\tilde{y}_{i}$ being the current prediction, $y_{i}$ being the current true value, and $N$ being the number of samples.
%
%
%

%
%
For all data analysis results, we apply 10-fold cross validation.
%
%
Based on the findings of related work, the following analyses focuses on using the \ac{RF} model for performing the data rate predictions. A deeper analysis about the performance of different machine learning models can be found in the in-depth study in \cite{Sliwa/Wietfeld/2019d}.

\subsection{Reference Discrete Event Simulation Setup in ns-3} \label{sec:reference_ns_3}

For comparison, a classic \ac{DES}-based setup is created using the \ac{LTE} framework \ac{LENA} \cite{Baldo/etal/2011a} for \ac{ns-3} \cite{Riley/Henderson/2010a}. 
All \acp{eNB} are positioned according to their corresponding real world locations. A summary of the simulation parameters is given in  Tab.~\ref{tab:parameters}.
However, since \ac{LENA} is not capable of representing the whole real world feature set -- \ac{CQI} and \ac{TA} are missing -- the prediction models need to be simplified.
As a result, the prediction performance is reduced: The average \ac{RMSE} is increased from $3.9$~MBit/s to $4.2$~MBit/s.

%
%
\newcommand{\entry}[2]{#1 & #2 \\}
\newcommand{\head}[2]{\toprule \entry{\textbf{#1}}{\textbf{#2}}\midrule}

\begin{table}[ht]
	\centering
	\vspace{-0.3cm}
	\caption{Parameters of the \ac{ns-3} Scenario}
	\vspace{-0.3cm}
	\begin{tabular}{ll}
		
		\head{Parameter}{Value}
		
		\entry{Carrier frequency}{\ac{eNB}-specific}
		\entry{Bandwidth}{20~MHz}
		\entry{Transmission power $P_{\text{TX}}$ (\acs{UE})}{23~dBm}
		\entry{Transmission power $P_{\text{TX}}$ (\acs{eNB})}{43~dBm}
		\entry{Channel model}{HybridBuildingsPropagationLossModel}
		\entry{Number of simulation runs}{30}
		
		\bottomrule
		
	\end{tabular}
	\label{tab:parameters}
	\vspace{-0.8cm}
\end{table}

\section{Results}  \label{sec:results}

In this section, the impact of using \ac{REM} for modeling radio channel conditions is evaluated. Afterwards, the proposed approach is validated against real world measurements and existing simulation methods.

%
%
\subsection{Radio Environmental Maps}

Due to the data aggregation performed within the \acp{REM}, the obtained values most likely differ from the individual measurements. Therefore, the impact of the aggregation granularity -- represented by the cell width $c$ -- on the prediction of individual indicators as well as on the overall data rate prediction is investigated.

Fig.~\ref{fig:cm_errors} shows the resulting lookup errors as \ac{RMSE} and \ac{MAE} functions for different network context indicators.
%
%
The highest accuracy is achieved for the smallest $c$ values where most cells only consist of a single measurement. However, in order to allow the usage of \acp{REM} within the simulation process, the cell size needs to be large enough to achieve sufficient coverage of the whole evaluation trajectory and minimize the lookup miss ratio which is shown in Fig.~\ref{fig:cm_errors}~(d). Remaining lookup misses can then be compensated by choosing the nearest neighboring cell.

As a direct consequence of these errors, also the machine learning based data rate prediction which uses the network context indicators as features is impacted by the chosen granularity. The resulting data rate prediction error in uplink and downlink direction is shown in Fig.~\ref{fig:cm_dataRate_errors}. Two different behaviors can be observed.
%
%
For $c\leq50~m$, a slight \emph{aggregation gain} is achieved. In this region, the channel coherence does not change significantly between different measurements in the same cell. Therefore, the  \ac{REM} acts like a filter which compensates short term fluctuations of the different measurements.
%
%
However, for $c>50~m$, the prediction accuracy is reduced for increasing $c$ values as the cell width is too large to represent the local radio propagation characteristics accurately.
%
%
This effect is more dominant in the uplink than in the downlink direction. As pointed out by the authors of \cite{Bui/etal/2017a}, the achievable downlink data rate is mainly determined by the resource competition between different cell users and less sensitive to radio propagation effects. 

%
%
\subsection{Validation} \label{sec:validation}

In the following, the proposed hybrid simulation method is compared to trace-based \ac{DDNS} according to \cite{Sliwa/Wietfeld/2019d}, \ac{ns-3}-based \ac{DES}, and real world measurements in the same scenario. For all simulation methods, the overall goal is to maximize the congruency with the real world measurements.

%
%
\begin{figure}[] 
	\centering
	\subfig{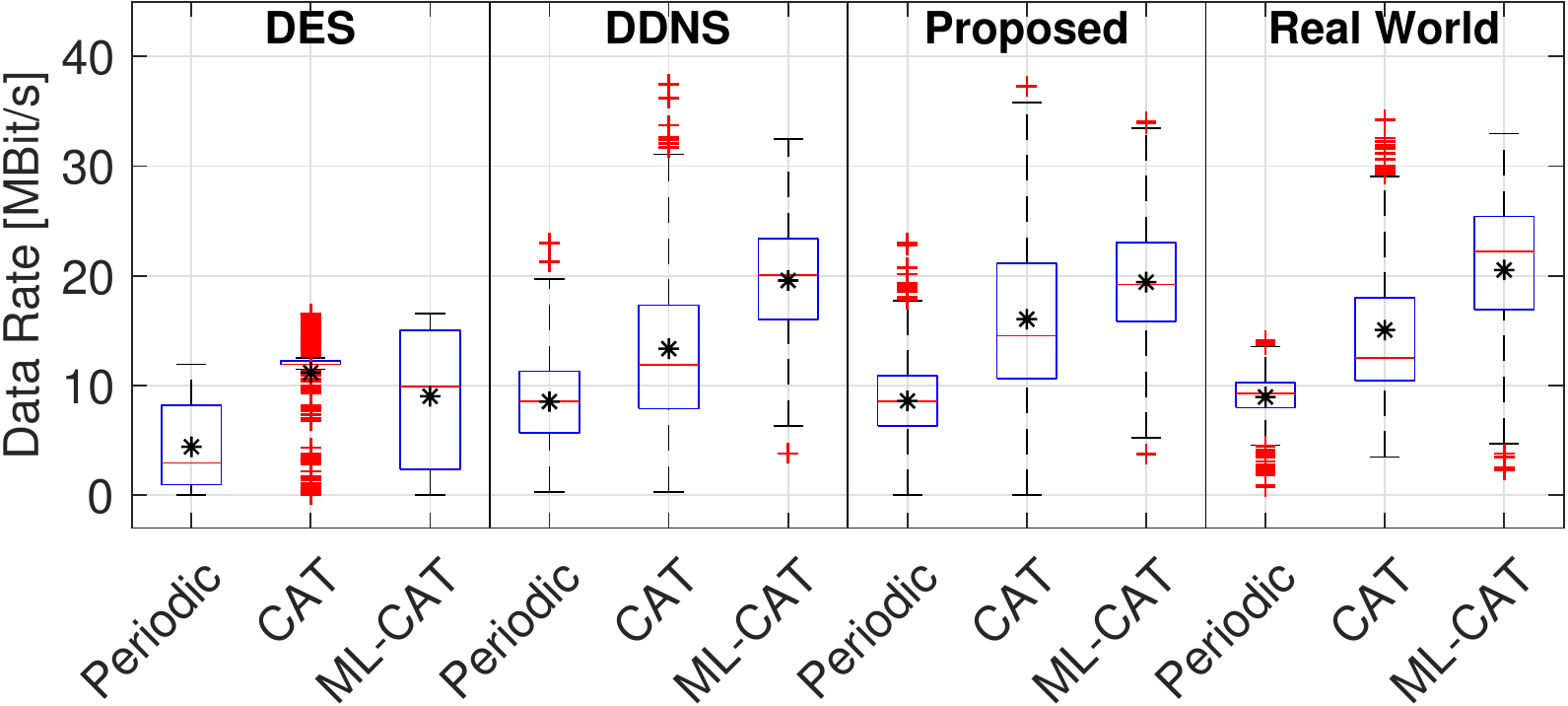}{.48}{Uplink}%
	\subfig{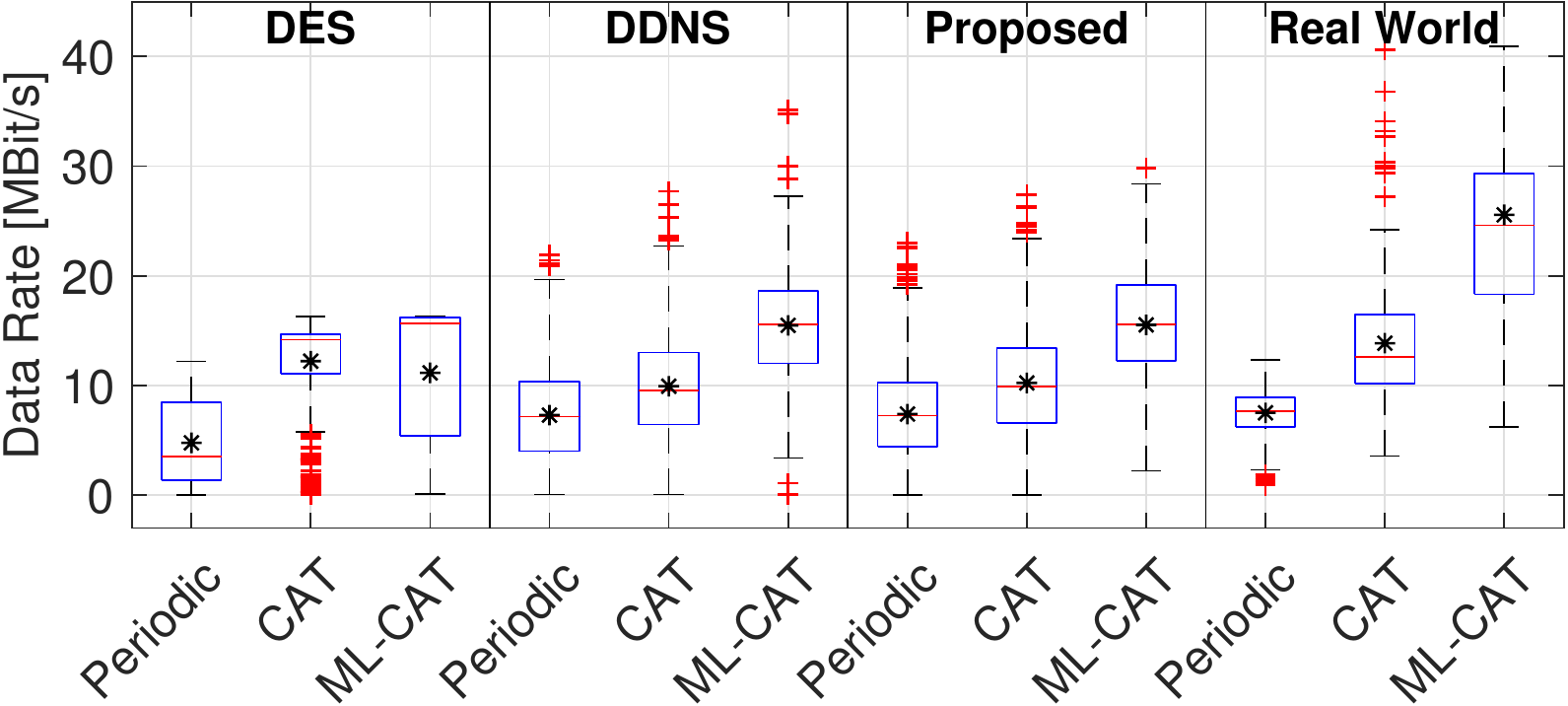}{.48}{Downlink}%
	
	\caption{Comparison of the resulting end-to-end behavior of different opportunistic transmission schemes with different evaluation methods.}
	\label{fig:boxplot_performance}
	\vspace{-0.5cm}
\end{figure}
The achieved data rate values for the different transmission schemes and performance evaluation methods are shown in Fig.~\ref{fig:boxplot_performance}. It can be seen that the highest overlap between real world measurements and corresponding simulated behaviors is achieved with the \ac{DDNS} method and the proposed hybrid simulation method.
%
%
The simulation-based representation of the real world behavior is more accurate in the uplink than in the downlink direction. As discussed in the previous analysis (see Fig.~\ref{fig:cm_dataRate_errors}), the machine learning models work more precise in the uplink direction. Here, the end-to-end behavior is more determined by channel related effects which are well covered by the utilized feature set.
As analyzed in \cite{Sliwa/etal/2020b}, the downlink data rate prediction accuracy could be significantly improved through consideration of load-depending features such as the number of active users and the amount of occupied \acp{PRB}. However, as the \acp{UE} are not aware of these indicators, it would be required to implement a \emph{cooperative} prediction approach where the \acp{eNB} actively distribute this information via control channel announcements.
%
%
In contrast to the data-driven approaches, the modeling accuracy of the \ac{DES} setup is significantly lower. Even more problematic, the massive improvements of the \ac{ML-CAT} method over the \ac{CAT} method are not represented at all: If the simulation-based performance analysis was used to make a decision for one or the other opportunistic data transmission method, the \ac{ns-3}-based approach would likely lead to a wrong decision.
As all \ac{CAT}-based methods rely on detecting and exploiting \emph{connectivity hotspots}, they are highly sensitive to the channel conditions. However, the stochastic channel models fail to represent the real world network behavior in the concrete evaluation scenario.
In addition, the need to simplify the prediction model for \ac{ML-CAT} (see Sec.~\ref{sec:reference_ns_3}) due to missing features results in a reduction of the accuracy. 
%
%
\begin{figure}[] 
	\centering
	\vspace{-0.4cm}
	\subfig{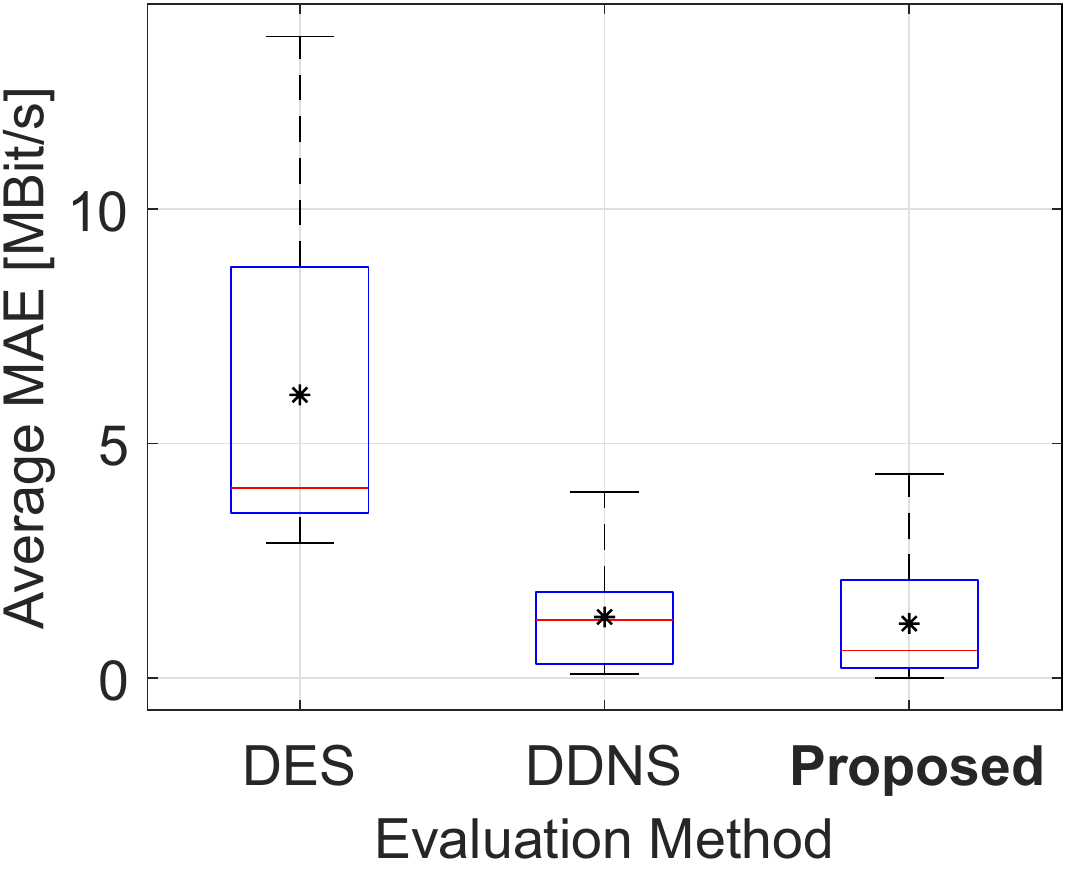}{0.24}{Uplink}%
	\subfig{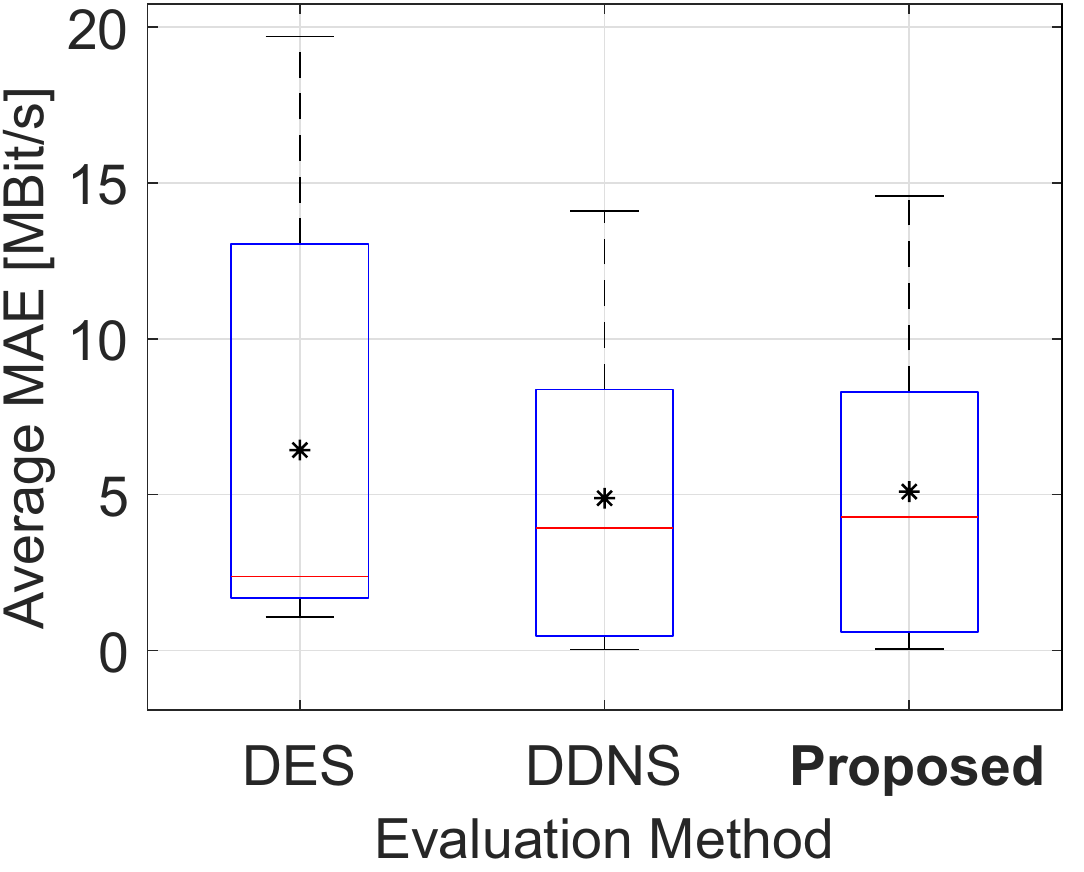}{0.24}{Downlink}%
	
	\caption{Relative aggregated modeling error for all considered simulation approaches.}
	\label{fig:total_accuracy}
\end{figure}
The aggregated modeling accuracy for all methods is shown in Fig.~\ref{fig:total_accuracy}.

%
%
\subsection{Computational Efficiency}

In addition to the achievable modeling accuracy, the required time to perform extensive simulation studies is another crucial factor that influences the choice of methods for the performance analysis.
%
%
\begin{figure}[]  	
	\vspace{0cm}
	\centering		  
	\includegraphics[width=1.0\columnwidth]{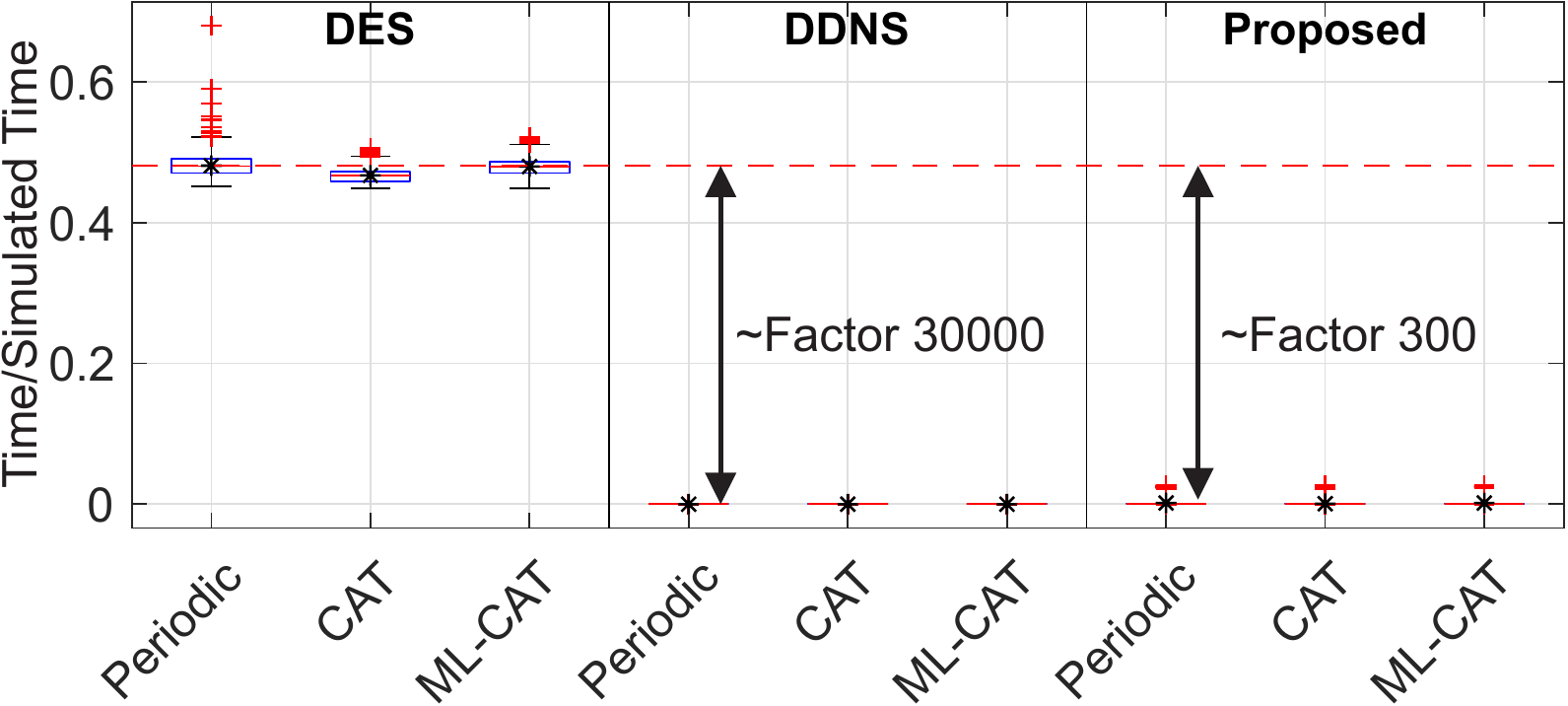}
	\caption{Comparison of the computational efficiency for the considered performance analysis methods.}
	\label{fig:evaluation_time}
	\vspace{-0.5cm}	
\end{figure}
A comparison of the computational efficiency of the considered methods is shown in Fig.~\ref{fig:evaluation_time}.
The highest computational efficiency is achieved with the pure \ac{DDNS} method which relies on context trace analysis. In comparison to the latter, the proposed hybrid method is impacted by the model-based mobility simulation (e.g., online routing) which reduces the simulation speed by a factor of 100. Still, it is able to benefit from the massive computational efficiency of the machine learning-based network simulation. The classical \ac{DES} approach represented by \ac{ns-3} has the lowest computational efficiency as it requires to explicitly model communicating entities as well as their protocol stacks.
For all methods, there are only marginal differences in the computation times of the different transmission schemes.
\section{Conclusion}

%
%
In this paper, we presented a hybrid approach for simulating the end-to-end performance of vehicular communication systems which brings together model-based mobility simulation, multi-dimensional \acp{REM}, and data-driven network simulation.
In constrast to existing methods that focus on modeling communicating entities and their corresponding protocol stacks, we utilize a combination of machine learning methods to model the end-to-end behavior of a target \ac{KPI}.
%
%
In a comprehensive validation campaign, the proposed method was able to mimic the real world behaviors of different opportunistic data transfer methods more accurately than a reference simulation setup in \ac{ns-3}. Moreover, the machine learning-enabled approach achieved a massively higher computational efficiency than classical system-level network simulation.
%
%
As the achievable accuracy of \ac{DDNS}-enabled simulation approaches is bound by the accuracy of the applied machine learning models, future work will focus on optimizing the latter, e.g., through application of cooperative prediction methods.
\ifdoubleblind

\else
\section*{Acknowledgment}

Part of the work on this paper has been supported by the German Federal Ministry of Education and Research (BMBF) in the project A-DRZ (13N14857) as well as by the Deutsche Forschungsgemeinschaft (DFG) within the Collaborative Research Center SFB 876 ``Providing Information by Resource-Constrained Analysis'', project B4.
\fi

\bibliographystyle{IEEEtran}
\bibliography{Bibliography}

\end{document}